\documentclass[aps,pre,showpacs,twocolumn,superscriptaddress]{revtex4}
\usepackage{epsfig}
\newcommand{\be}{\begin{equation}}
\newcommand{\ee}{\end{equation}}
\newcommand{\bea}{\begin{eqnarray}}
\newcommand{\eea}{\end{eqnarray}}

\newcommand{\ecoli}{{\it Escherichia coli }}
\newcommand{\colp}{{\it E.~coli}}
\newcommand{\coli}{{\it E.~coli }}

\begin{document}

\title{Graph animals, subgraph sampling and motif search in large networks}

\author{Kim Baskerville} \affiliation{Perimeter Institute for
  Theoretical Physics, Waterloo, Canada N2L 2Y5} 
\author{Peter Grassberger} \affiliation{Complexity Science Group,
  University of Calgary, Calgary, Canada} \affiliation{Institute for
  Biocomplexity and Informatics, University of Calgary, Calgary,
  Canada} 
\author{Maya Paczuski} \affiliation{Complexity Science Group,
  University of Calgary, Calgary, Canada}

\date{\today}

\begin{abstract}
  We generalize a sampling algorithm for lattice animals (connected
  clusters on a regular lattice) to a Monte Carlo algorithm for `graph
  animals', i.e. connected subgraphs in arbitrary networks. As with
  the algorithm in [N. Kashtan {\it et al.}, Bioinformatics {\bf 20},
  1746 (2004)], it provides a weighted sample, but the computation of
  the weights is much faster (linear in the size of subgraphs, instead
  of super-exponential). This allows subgraphs with up to ten or more
  nodes to be sampled with very high statistics, from arbitrarily
  large networks. Using this together with a heuristic algorithm for
  rapidly classifying isomorphic graphs, we present results for two
  protein interaction networks obtained using the TAP high throughput 
  method: one of \ecoli with 230 nodes and 695
  links, and one for yeast ({\it Saccharomyces cerevisiae}) with
  roughly ten times more nodes and links. We find in both cases that
  most connected subgraphs are strong motifs ($Z$-scores $>10$) or
  anti-motifs ($Z$-scores $<-10$) when the null model is the ensemble of
  networks with fixed degree sequence. Strong differences appear
  between the two networks, with dominant motifs in \coli being
  (nearly) bipartite graphs and having many pairs of nodes which
  connect to the same neighbors, while dominant motifs in yeast tend
  towards completeness or contain large cliques. We also explore a
  number of methods that do not rely on measurements of $Z$-scores or
  comparisons with null models. For instance, we discuss the influence
  of specific complexes like the 26S proteasome in yeast, where a
  small number of complexes dominate the $k$-cores with large $k$ and
  have a decisive effect on the strongest motifs with 6 to 8 nodes. We
  also present Zipf plots of counts versus rank. They show broad
  distributions that are not power laws, in contrast to the case when
  disconnected subgraphs are included.
\end{abstract}


\maketitle

\section{Introduction}

Recently, there has been an increased interest in complex networks,
partly triggered by the observation that naturally occurring networks
tend to have fat-tailed or even power law degree distributions
\cite{faloutsos,barabasi}. Thus real-world networks tend to be very
different from the completely random Erd\"os-Renyi \cite{bollobas}
networks that have been much studied by mathematicians, and which give
Poissonian degree distributions.  In addition, most networks have
further significant properties that arise either from functional
constraints, from the way they have grown (fat tails, e.g., are 
naturally explained by preferential attachment), or for other reasons.
As a consequence, a large number of statistical indicators have been
proposed to distinguish between networks with different functionality
(neural networks, protein transcription networks, social networks,
chip layouts, etc.) and between networks which were specially designed
or which have grown spontaneously (such as, e.g. the world wide web),
under more or less strong evolutionary pressure. These observables
include various centrality measures \cite{newman_SIAM}, assortativity
(the tendency of nodes with similar degree to link preferentially)
\cite{newman_SIAM}, clustering \cite{watts-stro,newman_clust},
different notions of modularity \cite{barabasi,ravasz,girvan,ziv,rosvall}, 
properties of loop statistics \cite{stadler},
the small world property (i.e., slow increase of the effective
diameter of the network with the number of nodes) \cite{milgram},
bipartivity (the prevalence of even-sized closed walks over closed
walks with an odd number of steps) \cite{estrada}, and others.

The frequency of specific subgraphs form a particular class of
indicators. Subgraphs that occur more frequently than expected are
referred to as motifs, while those occurring less frequently are
anti-motifs \cite{milo,shen-orr,vasquez,kashtan}. Typically, motif
search requires a null model for deciding when a subgraph is over- 
or under-abundant. The most popular null model so far has been the 
ensemble of all random graphs with the same degree sequence. This
popularity is largely due to the fact that it can be simulated easily
by means of the so-called `rewiring algorithm' \cite{besag,maslov}.
As we shall see, however, in the present analysis its value is severely 
limited, because it gives predictions
that are too far from those actually observed. Other null models that
retain more properties of the original network have been suggested
\cite{milo,mahadevan}, but have received much less attention. Analytic
approaches to null models are discussed in
Refs.~\cite{newman_park1,newman_park2,foster}.

\subsection{Motifs and the Search for Structure}

Up to now, motif search has been mainly restricted to small motifs,
typically with three or four nodes. Certain specific classes of 
larger subgraphs have been examined in
Refs.~\cite{class1,kashtan2004b,vasquez}. With the exception of
Ref.~\cite{baskerville}, few systematic attempts have been made to 
learn about significant structures at larger scale, by counting all 
possible subgraphs (for a different approach to the discovery of 
structure than discussed here see the work on inference of 
hierarchy in Ref.~\cite{clauset}).

One reason for this is that the number of non-isomorphic (i.e.
structurally different) subgraphs in any but the most trivial networks
increases extremely fast (super-exponentially) with their size. For
instance, the number of different undirected graphs with 11 nodes is
$\approx 10^9$~\cite{briggs}. Thus exhaustive studies of all possible
subgraphs with $>10$ nodes becomes virtually impossible with
present-day computers. But just because of this inflationary growth,
counts at intermediate sizes contain an enormous amount of potentially
useful information. Another obstacle is the notorious graph
isomorphism problem \cite{kobler,faulon}, which is in the NP class
(though probably not NP complete \cite{toran}). Existing state of the
art programs for determining whether any two graphs are isomorphic
\cite{nauty} remain too slow for our purpose.  Instead, we shall use
heuristics based on graph invariants similar to those put forward in
Ref.~\cite{baskerville}, where intermediate size motifs and
anti-motifs in the protein interaction network of \ecoli were
detected.

The last problem when studying larger motifs, and the main one
addressed in the present work, is the difficulty of estimating how
often each possible subgraph appears in a large network, i.e. of
obtaining a `subgraph census'.  Most studies so far were based on
exact enumeration. In a network with $N$ nodes, there are ${N\choose
n}$ subgraphs of size $n$. With $N=500$ and $n=6$, say, this number
is $\approx 5\times 10^{11}$. In addition, most of the subgraphs
generated this way on a sparse network would be disconnected, while
connected subgraphs are of more intrinsic interest. Thus some
statistical sampling is needed. If one is willing to generate disconnected 
as well as connected subgraphs, then uniform sampling is simple: Just 
choose random $n$-tuples of nodes from the network \cite{baskerville}. 
Uniform sampling connected subgraphs is less trivial. To 
our knowledge, the only work which addressed this systematically was 
Kashtan {\it et al.} \cite{kashtan2004b} (for a less systematic 
approach, see also \cite{spirin}). There, a biased sampling 
algorithm was put forward.  While generating the subgraphs is fast,
computing the weight factor needed to correct for the bias is
$\exp[O(n)]$, making their algorithm inefficient for $n\ge 7$.

\subsection{Graph Animals}

In the present paper we exploit the fact that sampling connected
subgraphs of a finite graph resembles sampling connected clusters of
sites on a regular lattice.  The latter is called the {\it lattice
  animal} problem \cite{animals}, whence we propose to call the
subgraph counting problem that of {\it graph animals}. It is important
to recognize obvious differences between the two cases. In particular,
lattices are infinite and translationally invariant, while networks
are finite and heterogeneous (disordered). For lattice animals one
counts the number of configurations up to translations (i.e. per unit
cell of the lattice), while on a network the quantity of immediate
interest is the absolute number of occurrences of particular
subgraphs. Still, apart from these issues, the basic operations
involved in both cases coincide.

Algorithms for enumerating lattice animals exactly exist and have been
pushed to high efficiency \cite{jensen}, but are far from trivial
\cite{redner}. Due to disorder, we should expect the situation to be
even worse for graph animals. Algorithms for stochastic sampling of
lattice animals are divided into two groups: Markov chain Monte Carlo
(MCMC) algorithms take a connected cluster and randomly deform it
while preserving connectivity \cite{stauffer,dickman,pivot}, while
`sequential' sampling algorithms grow the cluster from scratch
\cite{leath,hsu,gfn1998,care}. Even for regular lattices, MCMC
algorithms seem less efficient than growth algorithms \cite{hsu}. For
networks, this difference should be even more pronounced, since MCMC algorithms
would dwell in certain parts of the network, and averaging over the
different parts costs additional time. Thus we shall in the following
concentrate only on growth algorithms.

All growth algorithms similar to those in
\cite{leath,hsu,gfn1998,care} produce unbiased samples of {\it
  percolation} clusters. As explained in Sec.~II, this means that they
sample clusters or subgraphs with non-uniform probability (for an
alternative algorithm, see \cite{Redner79}). Consequently, computing
graph animal statistics requires the computation of weights to be
assigned to the clusters, in order to correct for the bias. In
contrast to the algorithm in Ref.~\cite{kashtan2004b}, the correct
weights are easily and rapidly calculated in our graph animal
algorithm.  This is its main advantage.

\subsection{Summary}

In Sec.~\ref{alg} we present the graph animal algorithm in detail. The 
method used to handle graph isomorphism is briefly reviewed in Sec.~III.
Extensive tests, mostly with two protein interaction networks, one for
\coli with 230 nodes and 695 links~\cite{ecoli}, and one for yeast
with 2559 nodes and 7031 links~\cite{yeast}, are presented in
Sec.~IV \cite{footnote}. Both networks were obtained using the TAP high 
throughput method. In particular, our algorithm involves as a
free parameter a percolation probability $p$. For optimal performance,
in lattice animals $p$ should be near the critical value where cluster
growth percolates~\cite{hsu}. We show how the performance for graph
animals depends on $p$, on the subgraph size $n$, and on other
parameters. In Sec.~V we use our sampling method to study these two
networks systematically. We verify that large subgraphs with high link
density are overwhelmingly strong motifs, while nearly all large
subgraphs with low link density are anti-motifs
\cite{vasquez,baskerville} -- although our data show much more
structure than suggested by the scaling arguments of \cite{vasquez}.
We also find striking differences in the strongest motifs for the two
networks.  Dominant motifs for the \coli network are either bipartite
or close to it (with many nodes sharing the same neighbors) while
`tadpoles' with bodies consisting of (almost) complete graphs dominate
for yeast.  Our conclusions and discussions of open problems are given
in Sec.~VI.

The present work only addresses undirected networks, but the graph
animal algorithm works without major changes also for directed
networks.  Due to the larger number of different directed subgraphs,
an exhaustive study of even moderately large subgraphs is much more
challenging~\cite{newpaper}.
% A first step in this direction will be given in

\section{The Algorithm}
\label{alg}
 
In this section we explain how our algorithm achieves uniform sampling
of connected subgraphs in undirected networks.  The graph animal
algorithm executes a generalization of the Leath algorithm for lattice
animals. The observation central to the work in
Refs.~\cite{leath,hsu,care} is that the animal and percolation
ensembles concern exactly the same clusters. The only difference
between the two ensembles is that clusters in the percolation ensemble
have different weights, while all clusters with the same number of nodes
(sites) have the same weight in the animal ensemble. We focus on site
percolation~\cite{stauffer-aharony}. Bond percolation could also be
used~\cite{hsu}, but this would be more complicated and is not
discussed here.

\subsection{Leath growth for graph animals}

For regular lattices and  undirected networks we use the following epidemic 
model for growing connected clusters of sites~\cite{leath}: \\
(1) Choose a number $p \in [0,1]$ and a maximal cluster size $n_{\rm max}$. 
Label all sites (nodes) as `unvisited'. \\
(2) Pick a random site (node)  $i_0$ as a {\it seed} for the cluster, so that 
the cluster consists initially of only this site; mark it as `visited'.\\
(3) Do the following step recursively, until all boundary sites of the 
cluster have been visited, or until the cluster consists of $n_{\rm max}$ 
sites, whichever comes first: (Note that a boundary site of a cluster $C$ is
a site which is  not in $C$, but which is connected to $C$ by
one or more edges). \\
(A) Choose one of the unvisited boundary sites of the present cluster, and 
mark it as visited;
(B) With probability $p$ join it to the cluster.\\
Once a boundary site has been visited, it cannot later join the cluster; it 
either joins the cluster when it is first visited (with probability $p$) or is 
permanently forbidden to join (with probability $1-p$).

The order in which the boundary (or `growth') sites are chosen
influences the efficiency of the algorithm, but this is irrelevant for
the present discussion. The growth algorithm can be seen as an
idealization of an epidemic process (`generalized' or SIR
epidemic~\cite{mollison, grass}) with three types of individuals
(Susceptible, Infected, Removed).  Starting with a single infected
individual with all others susceptible, the infected individual can
infect neighbours during a finite time span.  Everyone either gets
infected or doesn't at his/her first contact.  The latter are removed,
as are the infected ones after their recovery, and do not participate 
in the further spread of the epidemic.

Assume that for some fixed node $i_0$, a connected labeled subgraph $G^{\ell}$ 
exists, which contains $i_0$ and has $n<n_{\rm max}$ nodes and $b$ visited 
boundary nodes. The chance that precisely this particular labeled subgraph 
will be chosen using the algorithm is
\be 
    P_{G^{\ell}}(p;i_0) \equiv P_{nb}(p;i_0)= p^{n-1}(1-p)^b \;.  
\label{growth-prob} 
\ee
Since an independent decision is made at each boundary site, this is
indeed the probability for $n-1$ sites to be selected to join the
cluster, while $b$ sites are rejected.

Denote by $c(G^{\ell})$ the indicator function for the existence of $G^{\ell}$, 
i.e. $c(G^{\ell})=1$ if the subgraph exists in the network, and $c(G^{\ell})=0$ 
else. Furthermore, denote by $c(G^{\ell};i_0)$ the explicit indicator that 
$G^{\ell}$ exists and contains the node $i_0$. Then the total number of 
occurrences of the {\it unlabeled} subgraph $G$ is given by 
\be
    c_G = n^{-1} \sum_{i=1}^N c_{G,i} = n^{-1} \sum_{i_0=1}^N \sum_{G^{\ell}\sim 
G}c(G^{\ell};i_0),
\ee
where $c_{G,i}$ is the number of occurrences which contain node
$i$, and where the last sum runs over all labeled subgraphs
$G^{\ell}$ that are isomorphic to $G$. The factor $n^{-1}$ takes into
account that a subgraph with $n$ nodes is counted $n$ times.

If we repeat the epidemic process $M$ times, always starting at the
same node $i_0$, then the expected number of times $G^\ell$ occurs is
\be
   \langle m(G^\ell;p,i_0)\rangle = M c(G^\ell;i_0) P_{G^\ell}(p;i_0)\;.
\ee
Hence, an estimator for $c_{G,i}$ based on the actual counts
$m(G^\ell;P,i_0)$ after $M$ trials is
\be
   {\hat c}_{G,i_0}(M) = M^{-1} \sum_{G^{\ell}\sim G} m(G^\ell;p,i_0) 
[P_{G^\ell}(p;i_0)]^{-1} \;.
\ee
Here and in what follows carets always indicate estimators.

More generally, the starting nodes are chosen according to some
probability $Q_{i_0}$.  After $M>>1 $ trials in total, site $i_0$ will
have been used as starting point on average $Q_{i_0}M$ times. This gives then the
estimator for the total number of occurrences of $G$
\bea
        \label{simple_estimate}
   {\hat c}_G(M) & = & n^{-1} \sum_{i_0=1}^N {\hat c}_{G,i_0}(Q_{i_0}M) \\
         & = &(nM)^{-1} \sum_{i=1}^N Q_i^{-1} \sum_{G^{\ell}\sim G} m(G^{\ell};p,i) 
[P_{G^{\ell}}(p;i)]^{-1}. 
        \nonumber
\eea
It is simplest to take a uniform probability $Q_{i_0} = 1/N$. But a
better alternative is to choose each node with a probability
proportional to its degree, as nodes with larger degrees have more
connected subgraphs attached to them. This is accomplished by choosing
a link with uniform probability $1/L$, where $L$ is the total number
of links in the network, and then choosing one of the two ends of this
link at random. This gives
\be
   Q_i = (2L)^{-1}k_i.                                 \label{link-prob}
\ee

The algorithms of \cite{leath,care} are directly based on
Eq.~(5).  Their main drawback is that all
%Eq.~(\ref{simple_estimate}).  Their main drawback is that all
information from clusters which are still growing at size $n$ is not
used. Clusters whose growth had stopped at sizes $<n$ don't contribute
to ${\hat c}_G$ either, of course. Thus only those that stop growing
exactly at size $n$ are used in Eq.~(5). This
%exactly at size $n$ are used in Eq.~(\ref{simple_estimate}). This
requires, among other things, a careful choice of $p$: If $p$ is too
large, too many clusters survive past size $n$, while in the opposite
case too few reach this size at all. But even with the optimal choice
of $p$, most of the information is wasted.

\subsection{Improved Leath method}

The major improvement comes from the following observation \cite{hsu}:
Assume that a cluster has grown to size $n$, and among the $b$
boundary sites there are exactly $g$ which have not yet been tested
(`growth sites'). Thus growth has definitely stopped at $b-g$ already
visited boundary sites, while the growth on the remaining $g$ boundary
sites depends on future values of the random variable used to decide
whether they are going to be infected. With probability $(1-p)^g$ none
of them are susceptible, and the growth will stop at the present
cluster size $n$. Thus we can replace the counts $m(G^\ell;p,i_0)$ in
the estimator for $c_G$ by the counts of `unfinished' subgraphs,
provided we weigh each occurrence of a subgraph isomorphic to $G$ with
an additional weight factor $(1-p)^g$. Formally, this gives, with
uniform initial link selection (Eq.~(\ref{link-prob})), 
\bea
{\hat c}_G & = &{2L\over nM} \sum_{i=1}^N k_i^{-1} \sum_{G^{\ell}\sim G} 
p^{1-n}(1-p)^{g-b} \;\;\times  \nonumber \\
& & \times \;\; m_{\rm unfinished}(G^{\ell};p,i,g)\;.
    \label{estimate}
\eea
The quantity $m_{\rm unfinished}(G^{\ell};p,i,g)$ is the number of
epidemics (with parameter $p$) that start at node $i$, give a
labeled subgraph $G^{\ell}$ of infected nodes, and leave $g$
unvisited boundary nodes. The factor $p^{1-n}(1-p)^{g-b}$ has a simple
interpretation.  In analogy to Eq.~(\ref{growth-prob}) it is the
probability to grow a cluster with $n-1$ nodes in addition to the
start node, $g$ growth nodes, and $b-g$ blocked boundary nodes,
\be
   P_{nbg}(p;i_0) = p^{n-1}(1-p)^{b-g} \;.
\label{growth-prob-2}
\ee
Eq.~(\ref{estimate}) is the number of generated clusters, reweighted
with their inverse probabilities to be sampled, given they exist.  
It is the formula we use to estimate frequencies of occurrences of 
connected subgraphs in the protein interaction networks as discussed 
later in the text.

\subsection{Resampling}

In principle, Eq.~(\ref{estimate}) can be improved further.
Ref.~\cite{hsu} shows how to use the equivalents of
Eqs.(\ref{estimate},\ref{growth-prob-2}) for lattice animals as a
starting point for a re-sampling scheme. For completeness, re-sampling
for graph animals is briefly explained, even though it is not used in
this work.

For each cluster that is still growing a {\it fitness function} is defined as 
\be
   f_{nbg}(p) = p^{1-n}(1-p)^{-b} = [P_{nbg}(p;i_0)]^{-1}/ (1-p)^g.
\label{fitness}
\ee
Clusters with too small fitness are killed, while clusters with too
large fitness are cloned, with both the fitness and the weight being
split evenly among the clones. The first factor in the fitness is just
proportional to the weight, while the second factor takes into account
that clusters with larger $g$ have more possibilities to continue
their growth, and thus should be more `valuable'. The precise form of
Eq.(\ref{fitness}) is purely heuristic, but was found to be near
optimal in fairly extensive tests.

This resampling scheme was found to be essential, if one wants to
sample clusters of sizes $n>100$. In \cite{hsu}, the emphasis was on
very large clusters (several thousand sites), and thus resampling was
a necessity. Here, in contrast, we concentrate on subgraphs with
$\approx 10$ nodes or less, and stick to the simpler scheme without
resampling.  With respect to graph animals, we point out that optimal
fitness thresholds for pruning and cloning depend in a irregular
network on the start node, $i_0$, and have to be learned for each
$i_0$ separately. Although a similar strategy achieves success for
dealing with self avoiding walks on random lattices~\cite{randomSAW},
this is much more time consuming than for regular lattices.

\subsection{Implementation details}

For fast data access, we used several redundant data structures. The
adjacency matrix was stored directly as a $N\times N$ matrix with
elements 0/1 and as a list of linked pairs $(i,j)$, i.e. as an array
of size $L\times 2$. The first is needed for fast checking of which
links are present in a subgraph, while the second is the format in
which the networks were downloaded from the web. Finally, for fast
neighbor searches, the links were also stored in the form of linked
lists. To test whether a site was visited during the growth of the
present (say $k$-th, $k=1\ldots M$) subgraph, an array {\sf s[i]} of
size $N$ and type {\sf unsigned int} was used, which was initiated as
{\sf s[i]=0, i = 0,...N-1}. Each time a site {\sl i} was visited, we
set {\sf s[i] = k}, and {\sf s[i] $<$ k} was used as indicator that
this site had not been visited during the growth of the present
cluster.

In Leath-type cluster growth, there are two popular variants. Untested
sites in the boundary can be written either into a first-in first-out
queue, or into a stack (first-in last-out queue). In was found in
\cite{hsu} that these two possibilities, whose efficiency is roughly
the same when Eq.~(5) is used, give vastly
%the same when Eq.(\ref{simple_estimate}) is used, give vastly
different efficiency with Eq.(\ref{estimate}), in particular (but not
only) in combination with resampling. In that case, the first-in
first-out queue gives much better results, and we use this method to 
get the numerical results shown later.

\section{Subgraph Classification}

After sampling a labelled subgraph $G_\ell$, one has to find its 
isomorphism class $G$ (i.e., $G_\ell\sim G$), by testing which 
of the representatives for isomorphism classes it can be mapped onto by
permuting the node labels.  State-of-the-art computer programs for
comparing two graphs, such as NAUTY~\cite{nauty}, proceed in two
steps. First, some invariants are calculated such as the number of
links, traces of various powers of the adjacency matrix, a sorted list
of node degrees, etc. In most cases, this shows that the two graphs
are not isomorphic (if any of these invariants disagree), but
obviously this does not resolve all cases. When ambiguities remain,
each graph is transformed into a standard form by a suitable
permutation, and the standard forms are compared. The standard form
is, of course, also a special invariant, so the distinction between
``invariants" and ``standard form" might seem arbitrary. It becomes
relevant in practice, since the user of the package can specify which
invariants (s)he deems relevant, while the calculation of the
standard form is at the core of the algorithm and cannot be changed.

It is mostly the second step in this scheme which is time limiting 
and which renders it useless for our purposes -- although some 
invariants suggested e.g. by NAUTY are also quite demanding in CPU 
time. Thus we skip the second step and only use invariants that are fast to compute.
All these invariants, except for the number $n$ of nodes and the number
$\ell$ of links in the subgraph, are combined into a single index $I$, which is
intended to be a good discriminator between all non-isomorphic subraphs
with the same $n$ and $\ell$. Whenever a new subgraph is found, the
triplet ($n, \ell , I$) is calculated and compared to triplets that
have already appeared.  If the triplet appeared previously, the
counter for this triplet is increased by 1; if not, a new counter is
initiated and set to 1.

Since no known invariant (other than standard form) can discriminate
between any two graphs, any method not using it is necessarily
heuristic. Some of the invariants we used are those defined in
Ref.~\cite{baskerville}. In addition, we use invariants based on
powers of the adjacency matrix and of its compliment. More precisely,
if $A_{ij}$ is the adjacency matrix of a subgraph, then we define its
complement by $B_{ij} = 1-A_{ij}$ for $i\neq j$ and $B_{ij} = 0 =
A_{ij}$ for $i=j$. Any trace of any product $A^{a_1} B^{b_1} A^{a_2}
\ldots$ is invariant, and can be computed quickly. The same is true
for the number of non-zero elements of any such product, and for the
sum of all its matrix elements. The index $I$ is then either a linear
combination or a product (taken modulo $2^{32}$) of these invariants.
The particular choices were {\it ad hoc} and there is no reason to
believe they are optimal; hence those details are not given here.

With the indices described in~\cite{baskerville}, all undirected
graphs of sizes $n\leq 8$ and all directed graphs with up to $5$ nodes
are correctly classified. In this work, a faster algorithm for
counting loops is used; hence loop counting is always included, in
contrast to the work of \cite{baskerville}. Index calculation based on
matrix products is even faster but less precise: only 11112 out of all
11117 non-isomorphic connected graphs with $n=8$ were 
distinguished, and for directed graphs with $n=5$ just 4 graphs out of
9608~\cite{integer-seq} were missed. For larger subgraphs we were not
able to test the quality of the indices systematically, but we can
cite some results for $n=9$. Using indices based on matrix products, 
we found 239846 different connected subgraphs with $n=9$ in the \coli 
protein interaction network~\cite{ecoli} and its rewirings. Given the 
fact that there are only 261080 different connected graphs with $n=9$ 
\cite{integer-seq}, that many of them might not appear in the \coli 
network, and that our sampling was not exhaustive, our graph 
classification method failed to distinguish at most 9\% of the 
non-isomorphic graphs -- and probably many fewer.

\section {Numerical Tests of the Sampling Algorithm}

To test the graph animal algorithm, we first sampled both $n=4$ and
$n=5$ subgraphs of the \coli network, as well as $n=4$ subgraphs of
the yeast network. In these cases exact counts are possible, and we
verified that the results from sampling agreed with results from exact
enumeration within the estimated (very small) errors. To obtain these results 
we used crude estimates for optimal $p$ values, namely $p=0.11$ for \coli
and $p=0.03$ for yeast. For larger subgraphs more precise estimates
for the optimal $p$ are required.

\subsection{Optimal values for $p$}

When $p$ is too small, only small clusters are regularly encountered.
If $p$ is too large, performance decreases because the weight factors
in Eq.~(\ref{estimate}) depend too strongly on the number of blocked
boundary sites, $b-g$. The latter varies from instance to instance,
and this can create huge fluctuations in the weights given to
individual subgraphs.

The networks we are interested in are sparse ($L/N \approx const \ll
N$) and approximately scale-free \cite{yeast}.  As a result, most
nodes have only a few links, but some `hubs' have very high degree. In
fact, the degrees of the strongest hubs may diverge in the limit
$N\to\infty$.  For such networks it is well-known that the threshold
for spreading of an infinite SIR epidemic is zero~\cite{pastor}. On
finite networks this means that one can create huge clusters even for
minute $p$, and this tendency increases as $N$ increases.  Thus, we
anticipate the optimal $p$ to be small, and to decrease noticeably in
going from the \coli ($N=230$) to the yeast network ($N=2559$). This
is, in fact, what we find.

\begin{figure}
  \begin{center}
   \psfig{file=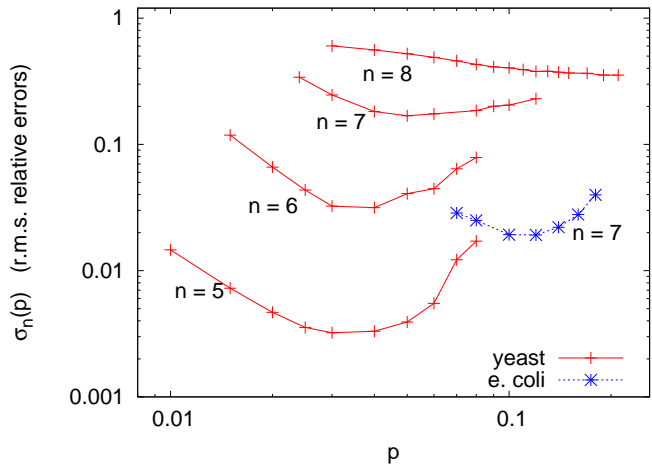,width=6.4cm,angle=270}
   \caption{(color online) Root mean square relative errors of connected
     subgraph counts, Eq.~(\ref{sigma}), for the yeast ($n=5$ to $8$)
     and \coli ($n=7$) networks. In most cases, clear minima indicate
     roughly the optimum value for $p$, with caveats as explained in
     the text. Each data point is based on $4\times 10^9$ generated
     subgraphs. Smaller values of $\sigma_n(p)$ indicate that the 
     census for subgraphs with $n$ nodes is on average more precise.}
\label{count-errors.fig}
\end{center}
\end{figure}

As a first test, we compute the root mean square relative errors of
the subgraph counts, averaged over all subgraphs of fixed size $n$.
Let $\gamma_n$ be the number of different subgraphs of size $n$ found,
and let $\Delta c_G$ be the error of the count for subgraph $G$. These
errors were estimated by dividing the set of $M$ independent samples
into bins, and estimating the fluctuations from bin to bin. Then
\be
   \sigma_n(p) = \left[{1\over \gamma_n} \sum_{j=1}^{\gamma_n}
           (\Delta c_{G_j}/{\hat c_{G_j}})^2\right]^{1/2}.
\label{sigma}
\ee
Smaller values of $\sigma_n(p)$ indicate that the subgraph census is
on average more precise.
Fig.~\ref{count-errors.fig} shows results for the yeast network, with
various values of $p$ and $n$. Also shown are data for the \coli
network, for $n=7$. Each simulation used for this figure (i.e., each
data point) involved $M = 4\times 10^9$ generated clusters. Our first
observation is that the results for \coli are much more precise than
those for yeast.  This is mainly due to smaller hubs ($k_{\rm
  max}^{\rm e. coli} = 36$, while $k_{\rm max}^{\rm yeast} = 141$), so
that much larger $p$ values~\cite{footnote2} could be used. Also in
all other aspects, our algorithm worked much better for the \coli
network than for yeast.  Therefore we exhibit in the rest of this
section only results for yeast, implying that whenever a test was
positive for yeast, an analogous test had been made for \coli with at
least as good results.

Even with the large sample sizes used in Fig.~\ref{count-errors.fig},
many $n=8$ subgraphs were found only once (in which case we set
$\Delta c_{G_j} /{\hat c}_{G_j}=1$), which explains the high values of
$\sigma_8(p)$. This is also why we do not show any data for $n>8$ in
Fig.~\ref{count-errors.fig}.  The relative error $\sigma_n(p)$ for
each $n<8$ shows a broad minimum as a function of $p$. The increase in
$\sigma_n(p)$ at small $p$ is because of the paucity of different
graphs being generated. This effect grows when $n$ increases, explaining why
the minimum shifts to the right with increasing $n$. The increase of
$\sigma_n(p)$ for large $p$, in contrast, comes from large
fluctuations of weights for individual sampled graphs.  When $p$ is
large, the factor $(1-p)^{b-g}$ in Eq.(\ref{estimate}) can also be
large, particularly in the presence of strong hubs.

Unfortunately, if a subgraph is found only once, it is impossible to
decide whether or not the frequency estimate is reliable. Even for
strong outliers, when the frequency estimate is far too large, the
formal error estimate cannot be larger than $\Delta c_G = O({\hat
  c}_G)$. This underestimates the true statistical errors and is
partially responsible for the fact that the curve for $n=8$ in
Fig.~\ref{count-errors.fig} does not increase at large $p$
\cite{footnote3}.

\begin{figure}
  \begin{center}
   \psfig{file=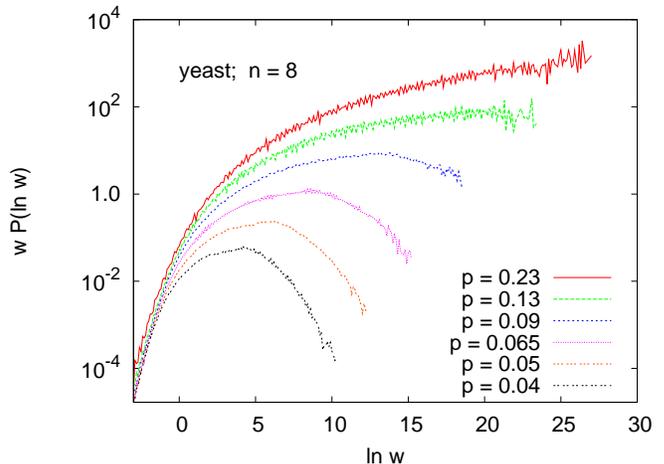,width=6.3cm,angle=270}
   \caption{(color online) Histograms of $wP(\ln w) = w^2P(w)$ for
     connected $n=8$ subgraphs of the yeast network. Each curve
     corresponds to one run ($4\times 10^9$ generated subgraphs)
     with fixed value of $p$. Results are the more reliable,
     the further to the left is the maximum of the curve and the
     faster is the decrease of its tail at large $w$.}
\label{weight-hist.fig}
\end{center}
\end{figure}

A more direct understanding of the decreasing performance at large
$p$ comes from histograms of the (logarithms of) weight factors. Such 
histograms, for $n=8$ subgraphs in the yeast network, are shown in 
Fig.~\ref{weight-hist.fig}. From the results in Section~\ref{alg}
\be
   w = {2L\over nMk}p^{1-n}(1-p)^{g-b}
\ee 
is the weight for a subgraph with $n$ nodes, $b$ boundary nodes, and
$g$ growth nodes. 
%$P(\ln w) = wP(w)$ is the probability distribution function of $\ln w$. 
The algorithm produces reliable estimates if $P(w)$ decreases for 
large $w$ faster than $1/w^2$, since averages
(which are weighted by $w$) are then dominated by subgraphs that are
well sampled. If, in contrast, $P(w)$ decreases more slowly, then the
tail of the distribution dominates, and the results cannot be taken at
face value \cite{grass-PERM}. We observe from
Fig.~\ref{weight-hist.fig} that the data for $n=8$ is indeed reliable
for $p<0.07$ only. The curve for $p=0.09$ in
Fig.~\ref{weight-hist.fig} also bends over at very large values of
$w$, indicating that even for this $p$ our estimates should finally be
reliable, when the sample sizes become sufficiently large.  But this
would require extremely large sample sizes.

\begin{figure}
  \begin{center}
   \psfig{file=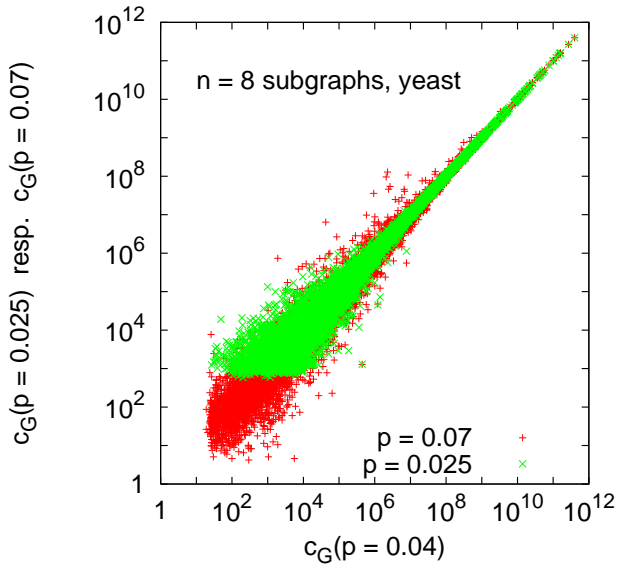,width=7.6cm,angle=270}
   \caption{(color online) Scatter plots of ${\hat c}_G(p=0.025)$ and
     ${\hat c}_G(p=0.07)$ against ${\hat c}_G(p=0.04)$ for connected $n=8$
     subgraphs of the yeast network. The clustering of the data along
     the diagonal indicates the basic reliability of the estimates,
     independent of the precise choice of $p$. Sample sizes were $4\times
     10^{10}$ for $p=0.04$, $2.4\times 10^{10}$ for $p=0.025$, and
     $8\times 10^9$ for $p=0.07$. The latter two correspond to roughly
     the same CPU time.}
\label{estimate-p.fig}
\end{center}
\end{figure}

As a last test we checked whether the estimates $\hat{c}_G$ are
independent of $p$ as they should be. Fig.~\ref{estimate-p.fig} shows
the estimates obtained for $n=8$ subgraphs in the yeast network with
$p=0.025$ and $p=0.07$ against those obtained with $p=0.04$.  Clearly,
the data cluster along the diagonal -- showing that the estimates are
basically correct. They scatter more when the counts are lower (i.e.
in the lower left corner of the plot).  The asymmetries in that region
result from the fact that rarely occurring subgraphs are completely
missed for $p=0.04$ and even more so for $p=0.025$, cutting off
thereby the distributions at small ${\hat c}_G$. For larger counts,
the estimates for $p=0.025$ are more precise than those for $p=0.07$.
The latter show high weight ``glitches" arising from the tail of
$P(w)$ discussed earlier in this section.

For increasing $p$, the numbers $m_G$ of generated subgraphs of type
$G$ increase of course (as the epidemic survives longer), so that
average weights, defined as $\langle w_G\rangle = {\hat c}_G M / m_G$,
decrease. But this decrease is not uniform for all $G$. Rather, it is
strongest for fully connected subgraphs ($\ell = n(n-1)/2$), and is
weakest for trees. For the yeast network and $n=8$, e.g., $\langle
w_G\rangle$ averaged over all trees decreases by a factor $\sim 18$
when $p$ increases from 0.025 to 0.085, while $\langle w_G\rangle$ averaged 
over all graphs with $\ell \ge 25$ decreases by a factor $\sim 1700$. Smaller
values of $\langle w_G\rangle$ are preferable, as they imply
smaller fluctuations.  Thus it would be most efficient to use larger
$p$ values for highly connected subgraphs, and smaller $p$ for
tree-like subgraphs.  Counting very highly connected subgraphs --
where every node has a degree in the subgraph $\geq k_0$, say -- is also made easier by
first reducing the network to its $k$-core with $k=k_0$, and then
sampling from the latter.

\section{Results}

\subsection{Characterization of the networks}

\begin{figure}
  \begin{center}
   \psfig{file=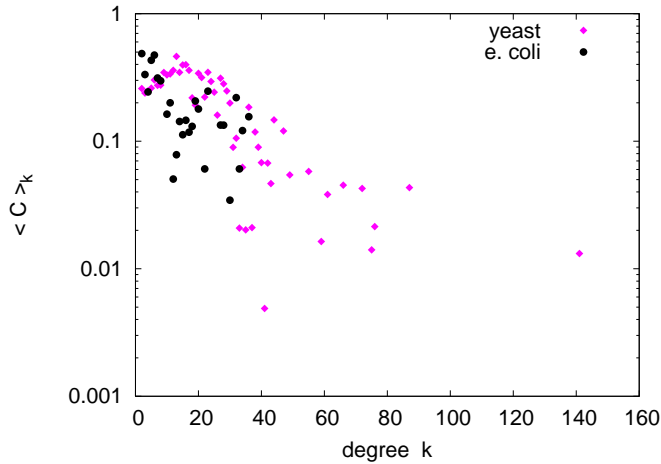,width=6.3cm,angle=270}
   \caption{(color online) Average clustering coefficients for nodes 
     with fixed degree $k$ plotted versus the degree, for
     the giant component of the yeast and \coli protein interaction
     networks. While the clustering coefficient decreases with
     $k$ for \colp, it attains a maximum at $k \approx 15$ for yeast.}
\label{clustering.fig}
\end{center}
\end{figure}

As already stated, both networks as we use them are fully connected 
\cite{footnote}.
The \coli network has 230 nodes and 695 links, while the yeast network
has 2559 nodes and 7031 links. Both networks show strong clustering,
as measured by the clustering coefficients \cite{watts-stro}
\be
   C_i = {2\over k_i(k_i-1)}\sum_{j<m} A_{jm}
\ee
where $k_i$ is the degree of node $i$ and the sum runs over all pairs
of nodes linked directly to $i$. In Fig.~\ref{clustering.fig} we show
averages of $C_i$ over all nodes with fixed degree $k$. We see that
$\langle C\rangle_k$ is quite large, but has a noticeably different
dependence on $k$ for the two networks.  While it decreases with $k$
for \colp, it attains a maximum at $k\approx 15$ for yeast.

\begin{figure}
  \begin{center}
   \psfig{file=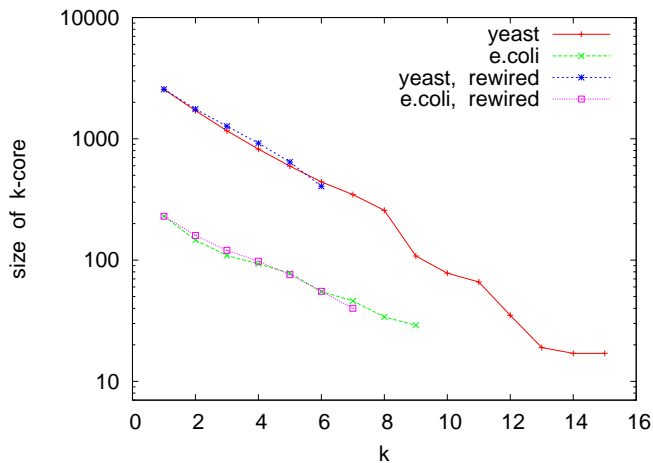,width=6.3cm,angle=270}
   \caption{(color online) Sizes of the $k-$cores for the two networks,
     plotted against $k$. Notice that the $k-$cores for yeast contain a 
     nearly fully connected cluster with 17 nodes. In addition to the 
     core sizes for the original networks, the figure also shows average
     core sizes for rewired networks as discussed in section V C.}
\label{k-cores.fig}
\end{center}
\end{figure} 

The unweighted average clustering ${\bar C} = N^{-1}\sum_{i=1}^N C_i$
is 0.1947 for yeast, and 0.2235 for \colp.  Due to the different
behavior of $\langle C\rangle_k$, the ranking is reversed for the
weighted averages
\be
    \langle C\rangle = {\sum_{i=1}^N C_i k_i(k_i-1) \over 
      \sum_{i=1}^N k_i(k_i-1)} = {3n_\Delta\over 3n_\Delta + n_\vee}, 
\ee 
where $n_\Delta$ is the number of fully connected triangles on the
network and $n_\vee$ is the number of triads with two links
(see~\cite{newman_clust} for a somewhat different formula).
Numerically, this gives $\langle C\rangle = 0.1948$ for yeast and
0.1552 for \colp. This can be understood as a consequence of the fact
that the relative frequency of fully connected triangles is higher in
yeast than in \colp: in yeast (\colp) there are 6969 (478) triangles
compared to 86291 (7805) triads with two links.

Associated with this difference are distinctions between the $k$-cores
\cite{seidman}
of the two networks. Fig.~\ref{k-cores.fig} shows the sizes of the
$k$-cores against $k$. We see that the yeast network contains
non-empty cores with $k$ up to 15. Moreover, the core with $k=15$ has
exactly 17 nodes. It is a nearly fully connected subgraph with just
one missing link. All 17 proteins in this core are parts of the 26S
proteasome which consists of 20 or 21 proteins \cite{mips,sgd}. All
these proteins presumably interact very strongly with each other. When the
interactions between the proteins within the 26S proteasome are taken
out (the corresponding elements of the adjacency matrix are set to
zero), the $k$-core with highest $k$ has $k=12$ and consists of 15
nodes. All its nodes correspond to proteins in the mediator complex of
RNA polymerase II \cite{mips}, which contains 20 proteins altogether.
After eliminating all interactions between these, two 11-cores with
respectively 13 and 14 nodes remain, the first corresponding to the
20S proteasome and the second corresponding to the RSC complex
\cite{mips}. Again these particular complexes have only a few more
proteins than those contained within their largest $k$-cores, so they
are very tightly bound together. All remaining complexes appear to be
more loosely bound, so that much of the strong larger scale clustering
in the yeast network (involving 7 - 10 nodes) can be traced to only a
few tightly bound complexes.  This has a big effect on the subgraph
counts, as we shall see.

\subsection{Trends in Subgraph counts}

\begin{figure}
  \begin{center}
   \psfig{file=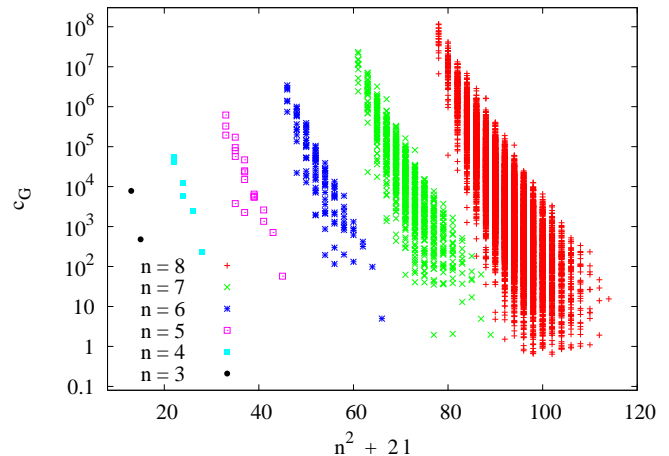,width=6.3cm,angle=270}
   \caption{(color online) Counts for connected subgraphs with fixed topology
     and with $n\le 8$ in the \coli network, plotted against $n^2
     +2\ell$. The variable $n^2 +2\ell$ is used to spread out the
     data, so that the dependence on both $n$ and $\ell$ (number of
     links) can be seen independently, without data points
     overlapping. For most of the points, the error bars are smaller
     than the sizes of the symbols.}
\label{ecoli-subgraphs.fig}
\end{center}
\end{figure}

\begin{figure}
  \begin{center}
   \psfig{file=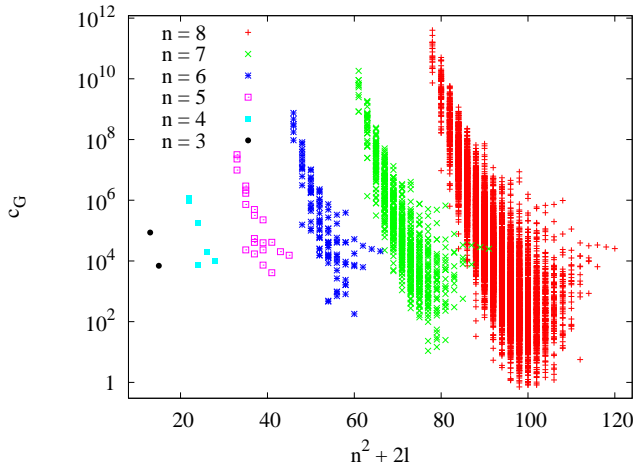,width=6.3cm,angle=270}
   \caption{(color online) Counts for subgraphs with fixed topology
     and with $n\le 8$ in the yeast network, plotted against
     $n^2+2\ell$ as in Fig.~\ref{ecoli-subgraphs.fig}.}
% \maya{Peter, do you know what the pdf of the counts looks like for a fixed n and l?}}
\label{yeast-subgraphs.fig}
\end{center}
\end{figure}

Subgraph counts ${\hat c}_G$ for the \ecoli and yeast networks, plotted
against $n^2 +2\ell$, are shown in Figs.~\ref{ecoli-subgraphs.fig} and
\ref{yeast-subgraphs.fig}. For large $n$ we see a very wide range,
with counts varying between 1 and $>10^8$.  In general, counts
decrease with increasing number of links, i.e. trees are most
frequent. This is a direct consequence of the fact that the networks
are sparse. Even when $n$ and $\ell$ are fixed, the counts $c_G$ can
range over six orders of magnitude (e.g. for yeast with $n=8$ and
$\ell=17$).

\begin{figure}
  \begin{center}
   \psfig{file=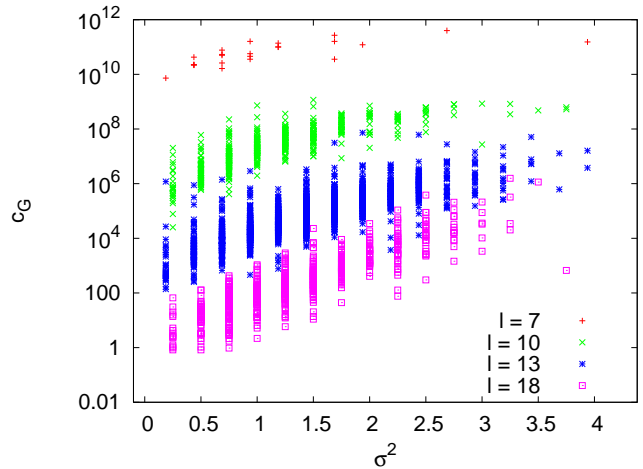,width=6.3cm,angle=270}
   \caption{(color online) Counts for $n=8$ subgraphs of the yeast
     network with $\ell = 7, 10,13,$ and $18$, plotted against the
     variance of the node degrees within the subgraphs, as given by 
     Eq.~\ref{var}. Zero variance means that all nodes
     have exactly the same degree, whereas a higher variance indicates
     that the nodes differ more widely. Typically, subgraphs
     with more variation in their nodes (and thus with larger $\sigma^2$) 
     have higher counts than those for which the degrees within the 
     subgraph are more uniform.}
\label{yeast-var.fig}
\end{center}
\end{figure}

For the yeast network, there are clear systematic trends for the
counts at fixed $n$ and $\ell$. The most frequent subgraphs are those
with strong heterogeneity, i.e. with a large variation of the degrees
(within the subgraph) of nodes, while the most rare are those with
minimal variation. Fig.~\ref{yeast-var.fig} shows the counts ${\hat c}_G$ for $n=8$
and with four different values of $\ell$ plotted against the variance
of the degrees of the nodes within the subgraph,
\be 
   \sigma^2 = {1\over n}\sum_{i=1}^n k_i^2 - [{1\over n}\sum_{i=1}^n k_i]^2.  
                   \label{var}
\ee
For all four curves we see a trend, where the count increases with
$\sigma$, but hardly any trend like this is seen for the \coli network
(data not shown). The effect seen in the yeast data is probably
related to the very strongly connected core in that network (see the
last subsection). As we shall also see later in subsection D,
subgraphs with high counts in yeast often have a tadpole form with a
highly connected body (which is part of one of the densely connected 
complexes discussed in the last subsection) and a short
tail attached to it. These cores may also be responsible for the main
difference between Figs.~\ref{ecoli-subgraphs.fig} and
\ref{yeast-subgraphs.fig}, namely the strong representation of very
highly connected (large $\ell$) subgraphs in the yeast network. Taking out all 
interactions within the 26S and 20S proteasomes, within the mediator
complex and within the RSC complex reduces substantially the counts for 
highly connected subgraphs. The count for the complete $n=7$ subgraph, e.g.,
is reduced in this way from $25,164\pm 68$ to $682\pm 23$. The removal 
of interactions within the 26S proteasome makes by far the biggest 
contribution.

\subsection{Zipf plots}

In~\cite{baskerville} it was found that ``Zipf plots" (subgraph counts
vs. rank) in the \coli network exhibit power law behavior, whose
origin is not yet understood. The essential difference between the
subgraph counts in~\cite{baskerville} and in the present paper is that
we sample only connected subgraphs, while {\it all} subgraphs with
given $n$ were ranked in~\cite{baskerville}.  Also, noting that
disconnected subgraphs are more likely to be sampled than connected
ones when picking nodes at random (due to the sparsity of the
networks), we can go to much higher ranks for the connected subgraphs.

\begin{figure}
 \begin{center}
  \psfig{file=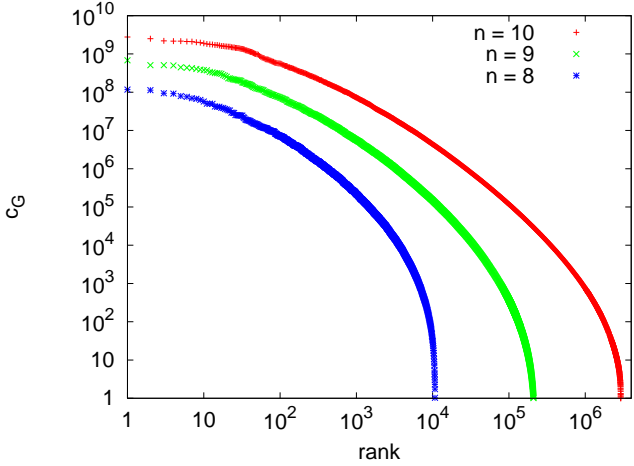,width=6.3cm,angle=270}
  \caption{(color online) ``Zipf" plots showing the counts for individual
    connected subgraphs with fixed $n$, plotted against their rank.
    Data are for the \coli network.}
\label{zipf.fig}
\end{center}
\end{figure}

Zipf plots for connected subgraphs in the \coli network are shown in
Fig.~\ref{zipf.fig}. Each curve is based on $4\times 10^9$ to
$10^{10}$ generated subgraphs. Each is strongly curved,
suggesting that there are no power laws -- at least for subgraph sizes
where we obtain reasonable statistics for the census. The curves
show less curvature for larger $n$, but this is a gradual effect. It
seems that the scaling behavior found in \cite{baskerville} was mainly
due to the presence of disconnected graphs, although it is not
immediately obvious why those should give scale-free statistics
either. In addition, the right hand tails of the Zipf plots in 
\cite{baskerville} were cut
off because of substantially lower statistics. In our case, apparently
sharp cutoffs in the counts are observed for ranks $\approx
1.08\times 10^4$ for $n=8$, $\approx 2.1\times 10^5$ for $n=9$, and
$\approx 2.9\times 10^6$ for $n=10$. For $n\leq 9$ these are close to
the total number of different connected subgraphs~\cite{briggs},
suggesting that we have fairly complete statistics. For $n=10$ the
cutoff is more affected by lack of statistics, but it is still within
a factor of four of the upper limit.

\subsection{Null model comparison and motifs}

One of the most striking results of \cite{baskerville} was that most
large subgraphs were either strong motifs or strong anti-motifs.
However, this finding was based on rather limited statistics and on a
single protein interaction network.  One of the purposes of the
present study is to test this and other results of \cite{baskerville}
with much higher statistics and for a larger network, the protein
interaction network of yeast.

\begin{figure}
  \begin{center}
   \psfig{file=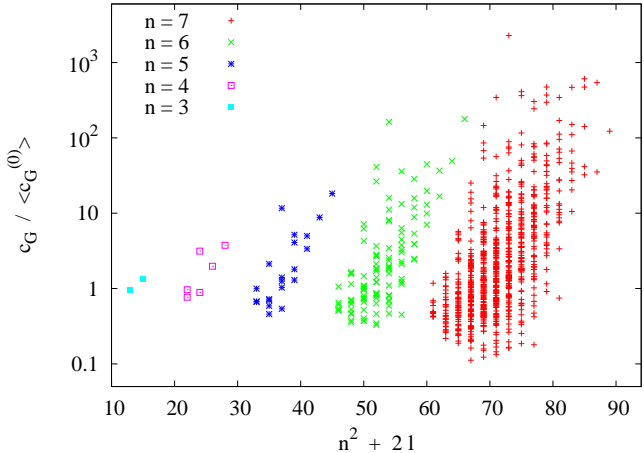,width=6.3cm,angle=270}
   \caption{(color online) Ratios between the count estimates
     $\hat{c}_G$ for connected subgraphs in the \coli
     network, and the corresponding average counts $\langle\hat{c}^{(0)}_G\rangle$
     in rewired networks. The data are plotted against $n^2+2\ell$,
     again to spread the points out conveniently. Most error bars are
     smaller than the symbols.}
\label{nullratio-ecoli.fig}
\end{center}
\end{figure}

\begin{figure}
  \begin{center}
   \psfig{file=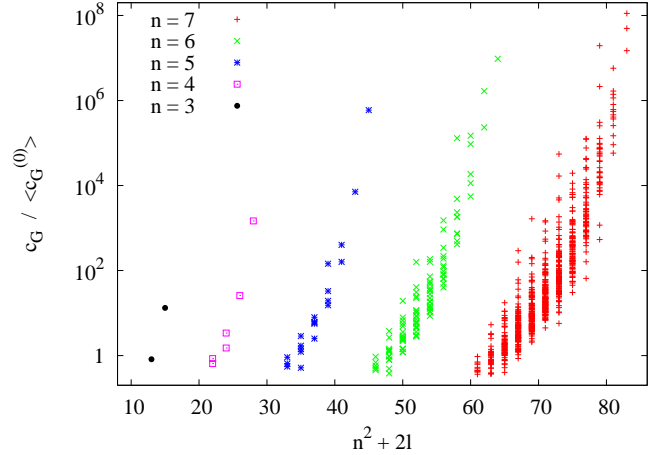,width=6.3cm,angle=270}
   \caption{(color online) Same as Fig.~\ref{nullratio-ecoli.fig}, but for 
     the yeast network.  Notice that most data points for large $n$ and
     $\ell$ are missing. Indeed, for $n=7$ all (!) data points with $\ell >
     16$ are missing, because no such subgraphs were found in the
     rewired ensemble.}
\label{nullratio-yeast.fig}
\end{center}
\end{figure}

To define a motif requires a null model. We take this to be the ensemble 
of networks with the same degree sequence, obtained by the rewiring 
method.  The average subgraph counts in the null ensemble are denoted 
as $\langle c_G^{(0)}\rangle$.  In Figs.~\ref{nullratio-ecoli.fig} and
\ref{nullratio-yeast.fig} we plot the ratios $c_G / \langle c_G^{(0)}\rangle$ 
against the variable $n^2+2\ell$ for each connected subgraph that was sampled 
both in the original graph and in at least one of the rewired graphs. 
The error bars, which include both
statistical errors from sampling and the ensemble fluctuations of the
null model estimated from several hundred rewired networks, are for
most points smaller than the symbols. A subgraph is a motif
(anti-motif), if this ratio is significantly larger (smaller) than 1.
Notice that motifs do not in general occur particularly
frequently in the original network. Even without rigorous estimates 
to estimate significance, it is clear that most densely connected 
subgraphs are motifs in the yeast network. The fact that trees or 
subgraphs with few loops tend to be anti-motifs might not be so evident 
from Fig.~\ref{nullratio-yeast.fig}, since the ratios for trees and
tree-like graphs are close to one. Thus we have to discuss
significance more formally.

\subsubsection{$Z$-scores}

Usually~\cite{baskerville}, the significance of a motif (or
anti-motif) is measured by its $Z$-score
\be 
   Z = {c_G - \langle c_G^{(0)}\rangle \over \sigma_G^{(0)}}\;,
              \label{Z}
\ee
where $\sigma_G^{(0)}$ is the standard deviation of $c_G$ within the null 
ensemble. A subgraph is a motif (anti-motif), if $Z \gg 1$ ($Z\ll -1$).

\begin{figure}
  \begin{center}
   \psfig{file=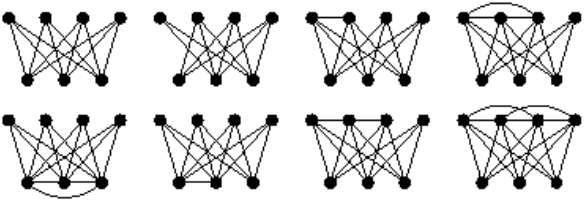,width=8.5cm,angle=0}
   \caption{The eight strongest motifs with $n=7$ in the \coli protein
     interaction network. These tend to be almost bipartite graphs,
     and many pairs of nodes are linked to the same set of neighbors.
     Their $Z$-scores, in order from left to right, first then second
     row, are: $2.9\times 10^4, 932, 885, 648, 595, 532, 516$ and
     377. Their estimated frequencies in the original \coli network 
     are, in the same order: $20936\pm 8,
     161521\pm 63, 8312\pm 5, 1331\pm 2, 838\pm 2, 5985 \pm 5, 5165\pm 4,$
     and $ 519\pm 1$.}
\label{fig:ecoli_motif}
\end{center}
\end{figure}

\begin{figure}
  \begin{center}
   \psfig{file=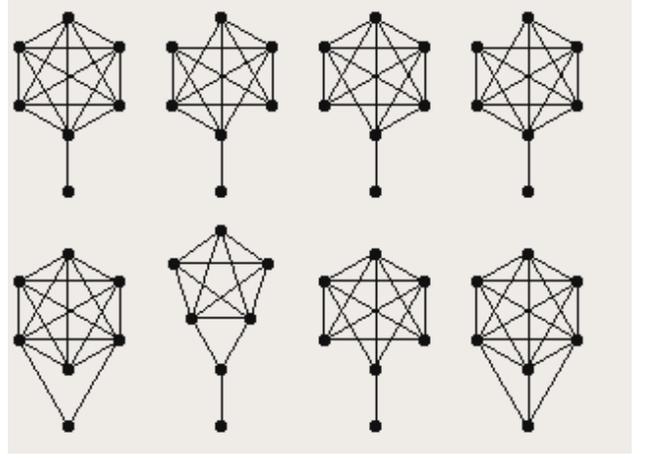,width=8.3cm,angle=0}
   \caption{Eight very strong motifs with $n=7$ for the yeast protein
     interaction network. These tend to be almost complete graphs with
     a single dangling node.  Four of these graphs were not seen in
     any realization of the null model, so only lower bounds on their 
     $Z$-scores can be given. From left to right, first then second
     row, the estimated $Z$-scores are: $>3\times 10^7, 9\times 10^5,
     >8\times 10^6, 5\times 10^5,>4\times 10^6,3\times 10^5,2.5\times 10^5$,
     and $>1.5\times 10^6$. Estimated frequencies are, in the same order:
     $6.68(1)\times 10^5, 9.27(5)\times 10^4, 1.76(1)\times 10^5, 4.84(1)\times 10^5, 
     7.78(2)\times 10^4, 3.13(6)\times 10^5, 1.38(1)\times 10^5$, and
     $3.35(1)\times 10^4$.}
\label{fig:yeast_motif}
\end{center}
\end{figure}

The eight strongest motifs with $n=7$ in the \coli network according
to this definition are shown in Fig.~\ref{fig:ecoli_motif}, together
with their $Z$-values. To name the strongest motifs in the yeast
network is less straight forward, since many subgraphs did not show
up in any rewired network at all. Assuming for those subgraphs 
$\sigma_G^{(0)} = \langle c_G^{(0)}\rangle = 0$ would give $Z=\infty$. 
Rough lower bounds on $Z$ are obtained for them by assuming that $\langle
c_G^{(0)}\rangle < 1/R$ and $\sigma_G^{(0)} < 1/\sqrt{R}$, where $R$
is the number of rewired networks that were sampled, giving $Z\geq c_G\sqrt{R}$. 
Some of the strongest motifs in the yeast network, together with their
estimated $Z$-scores, are shown in Fig.~\ref{fig:yeast_motif}. Note
that no $n=7$ graphs with $\ell>16$ were found in any of the realizations of
the null model, while they were all found in the real yeast network. Hence
these are all strong motifs. Those motifs in Fig.~\ref{fig:yeast_motif} for 
which only lower bounds for the $Z$-score are given are the most frequent in 
the real network, hence they have the highest lower bound. It was
pointed out in \cite{spirin,ispolatov} that cliques (complete subgraphs) 
are in general very strong motifs. In yeast, the $n=7$ clique (with
$\ell=21$) is indeed a very strong motif, but it does not have the largest
lower bound on the $Z$-score.  In comparison, anti-motifs have rather
modest $Z$-scores. The strongest anti-motif with $n=7$ has $Z=-32.9$
($Z=-24.7$) for \coli (yeast).

With $Z$-values up to $10^7$ and more, as in
Fig.~\ref{fig:yeast_motif}, the motivation for using $Z$-scores
becomes suspect. On the one hand, the null model is clearly unable
to describe the actual network, and has to be replaced by a more
refined null model. This will be done in a future paper
\cite{newpaper}. On the other hand, it suggests to use instead a
$Z$-score based on {\it logarithms} of counts,
\be 
Z_{\rm log} = {\log c_G - \langle \log c_G^{(0)}\rangle \over
  \sigma_{\log, G}^{(0)}}\;,
\label{Z_log}
\ee
where $\sigma_{\log, G}^{(0)}$ is the standard deviation of $\log
c_G^{(0)}$.  An advantage of Eq.(\ref{Z_log}) would be that it
suppresses $|Z|$ for motifs, but enhances $|Z|$ for anti-motifs.

In general, strong yeast motifs have a tadpole structure with a
complete or almost complete body, and a tail consisting of a few nodes
with low degree. This agrees nicely with our previous observation that
frequently occurring subgraphs in the yeast network have strong
heterogeneity in the degrees of their nodes.  In contrast, strong \coli 
motifs with not too many loops are all based on a 4-3 or 5-2 bipartite
structure. When the number of loops increases, strictly bipartite
structures are impossible, but the tendency towards these structures
is still observed. 

Whether we use $Z$-scores or the ratio $C_G/C_G^{(0)}$ to identify
motifs makes very little difference. Using either criterion, the
strengths of the strongest motifs skyrocket with subgraph size. This
is most dramatically apparent for the yeast network. Indeed,
correlations between $Z$-scores of individual graphs in the yeast and
\coli networks (data not shown) are much weaker than correlations
between count ratios. The latter are shown in Fig.~\ref{graph-r} for
$n=7$ subgraphs.

\subsubsection{Twinning versus Clustering}

Another characteristic feature of strong motifs in the \coli network is 
the tendency for `twin' nodes. We call two nodes in a subgraph twins if
they are connected to the same set of neighbours in the subgraph.
Otherwise said, nodes $i$ and $k$ are twins, iff the $i$-th and $k$-th
rows of the subgraph adjacency matrix are identical. Notice that twin
nodes can be created most naturally by duplicating genes. We
found that subgraphs with many pairs of twin nodes are in general also
motifs in the yeast network, but they do not stand out spectacularly
from the mass of other motifs. They could be the `genuine' motifs also
for yeast, but only a better null model where all subgraphs actually 
occur with reasonable frequency would be able to prove or disprove this.

\begin{figure}
  \begin{center}
\epsfig{file=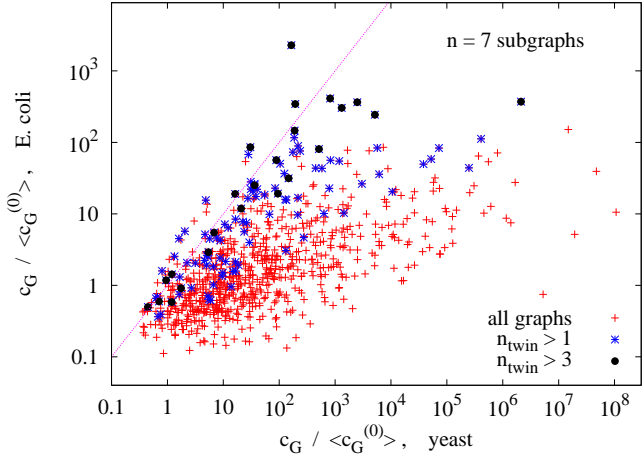, width=6.3cm, angle=270}
\caption{(color online) Count ratios $c_G /\langle c_G^{(0)}\rangle$ for individual
  subgraphs in the \coli network, plotted against the count ratio for
  the same subgraph in the yeast network. To highlight the dependence
  on the number of twin nodes in the subgraph, subgraphs with $n_{\rm
    twin}>1 \; (n_{\rm twin}> 3)$ are marked by asterisks
  (bullets). Whereas almost all ratios are much higher in the yeast
  network, this is noticeably less true for subgraphs containing
  more than three pairs of twin nodes. These tend to fall on the 
  diagonal indicated by the dashed line.}
\label{graph-r}
\end{center}
\end{figure}

In Fig.~\ref{graph-r} we also indicated the dependence on the number 
$n_{\rm twin}$ of pairs of twin nodes, by marking subgraphs with 
$n_{\rm twin}>3$ ($n_{\rm twin}> 1)$ by bullets (asterisks). We 
see that all strong motifs in \coli have multiple pairs of twin nodes.
These subgraphs tend to be also motifs of comparable strength in yeast
-- the bullets in Fig.~\ref{graph-r} tend to cluster on the diagonal
$[c_G /\langle c_G^{(0)}\rangle]_\coli = [c_G /\langle c_G^{(0)}\rangle]_{yeast}$. However, there
are even stronger motifs in yeast that have no twin nodes. These
graphs are typically much weaker motifs or not motifs at all in \colp.

As we have already indicated, many of the strong motifs in yeast seem to 
be related to a few densely connected complexes such as those discussed in
subsection A. They are either part of their cores, or they have most of
their nodes in the core, with one or two extra nodes forming the tail of
what looks like a tadpole. This effect is even more pronounced for 
$n=8$ subgraphs. For instance, the three most frequent subgraphs with 
$n=8$ and $ \ell = 17$ all contained a 6-clique and two nodes connected 
to it either in chain or in parallel. None of them occurred even in a 
single rewired network.

The situation is different for the \coli network. There, the three 
most frequent graphs with 8 nodes and 17 edges also have a tadpole 
structure, few twin nodes, and low bipartivity. But they are not very 
strong motifs since they occur also frequently in the rewired networks. 
The three strongest motifs with $n=8$ and $ \ell = 17$, in contrast, 
have many twin pairs and high bipartivity. They have slightly lower 
counts (by factors 2-4), but occur much more rarely in the rewired 
networks. 

\begin{figure}
  \begin{center}
  \epsfig{file=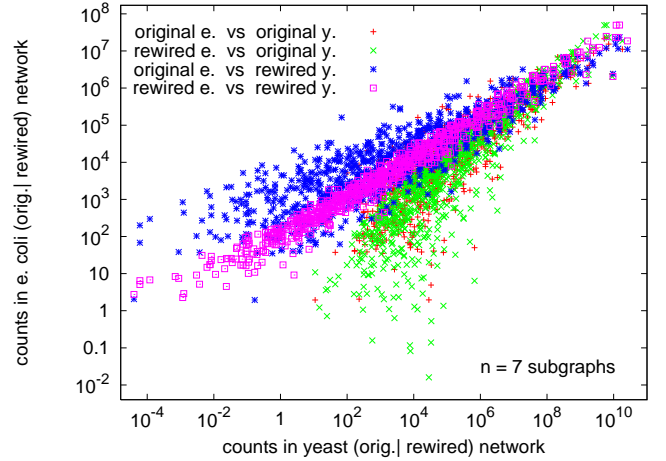, width=6.3cm, angle=270}
  \caption{(color online) Counts $c_G$ resp. $\langle
    c_G^{(0)}\rangle$ for individual subgraphs in the \ecoli network,
    plotted against counts for the same subgraph in the yeast network.
    It can be seen that the two rewired networks are much more similar
    (display higher correlation) than the original networks.}
  \label{graph-freq}
  \end{center}
\end{figure}

\subsubsection{Effects of Rewiring on Differences between Networks}

Finally, Fig.~\ref{graph-freq} shows counts for individual subgraphs
in the \coli network against counts for the same subgraph in yeast. 
This is done for all four combinations of original and
rewired networks. We see that the correlation is strongest when we
compare rewired networks of \coli to rewired networks of yeast. This
is not surprising. It means that a lack of correlations is mostly due
to special features of one network which are not shared by the other.
Rewiring eliminates most of these features. The other observation is
that rewiring in general reduces further the counts for subgraphs
which are already rare in the original networks. This is mainly due to
the fact that such subgraphs are relatively densely connected, and
appear in the original networks only because of the strong clustering.
This effect is more pronounced for yeast than for \colp, because it 
is more sparse and has more densely connected clusters/complexes.

\section{Discussion}

In this paper we have presented an algorithm for sampling connected
subgraphs uniformly from large networks. This algorithm is a
generalization of algorithms for sampling lattice animals, hence we
refer to it as a ``graph animal algorithm" and to the connected subgraphs 
as ``graph animals". It allowed us to obtain high statistics estimates of
subgraph censuses for two protein interaction networks. Although the
graph animal algorithm worked well in both cases, the analysis of the
smaller network (\colp) was much easier than that of the bigger (yeast). 
This was not so much because of the sheer size of the latter (the yeast 
network has about ten times more nodes and links than the \coli network), 
but was mainly caused by the existence of stronger hubs. Indeed, the
presence of hubs places a more stringent limitation on the method than
the size of the network.

One of the main results is that many subgraph frequency counts are
hugely different from those in the most popular null model, which is
the ensemble of networks with fixed degree sequence. Based on a
comparison with this null model, most subgraphs with size $\geq 6$ in
both networks would be very strong motifs or anti-motifs. This clearly
shows that alternative null models are needed which take clustering and
other effects into account.

While this was not very surprising (hints of it had been found in
previous analyses), a more surprising result is the fact that the
dominant motifs in the two protein interaction networks show very
different features. Most of these seem to be related to the densely
connected cores of a small number of complexes in the yeast network,
which have no parallels in the \coli network and which strongly affect
the subgraph census. Further studies are needed to disentangle
these effects from other -- possibly biologically more interesting -- 
effects.

Finally, a feature with likely biological significance is the dominance 
of subgraphs with many twin nodes. These are nodes which share the 
same list of linked neighbors within the subgraph. They correspond to 
proteins which interact with the same set of other proteins. The most 
natural explanation for them is gene duplication.  Connected to 
this is a preference for (approximately) bipartite subgraphs. These 
two features are very clearly seen in the \coli network, much less so 
in yeast. But it would be premature to conclude that gene duplication 
was evolutionary more important in \coli than in yeast. It is more 
likely that its effect is just masked in the yeast network by other 
effects, most probably by the densely connected complexes and other
clustering effects which do not show up to the same extent in \colp.

\begin{figure}
  \begin{center}
  \epsfig{file=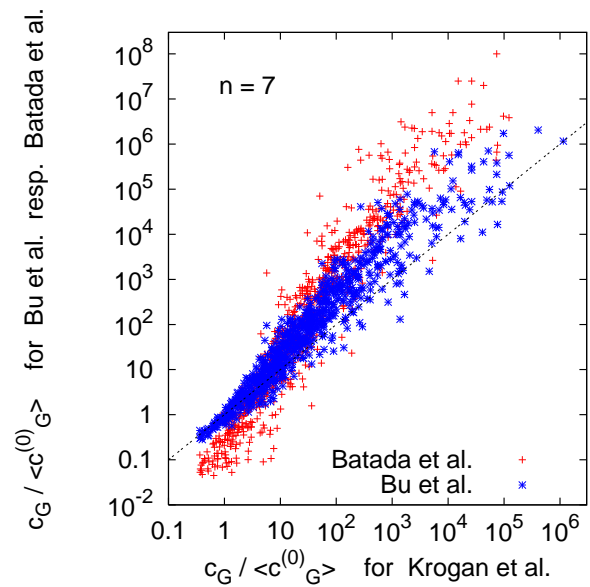, width=7.7cm, angle=270}
  \caption{(color online) Count ratios $c_G/\langle c_G^{(0)}\rangle$
    for individual subgraphs in the yeast networks of
    Refs.~\cite{bu,batada}, plotted against counts for the same
    subgraph in the network of \cite{yeast}. If all three networks
    were identical, all points should lie on the diagonal (indicated by
    the straight dashed line), whereas in fact systematic deviations are 
    observed.}
  \label{ratios-compare}
  \end{center}
\end{figure}

Up to now, we know very little about the biological significance of our 
findings. One main avenue of further work could be to relate our results 
on subgraph abundances in more detail to properties of the network that
are associated with biological function. Another important problem is the 
comparison between network reconstructions which supposedly describe 
the same or similar objects. There exist, e.g., a large number of 
published protein-protein interaction networks for yeast.
Some were obtained by means of different experimental techniques, either
with conventional or with high throughput methods, while others were 
obtained by comprehensive literature compilations. In a preliminary 
step, we compared three such networks: The network obtained by Krogan
{\it et al.}~\cite{yeast} that was studied above, a somewhat older 
network downloaded from~\cite{pajek} and attributed to 
Bu {\it et al.}~\cite{bu}, and the `high confidence' (HC) 
network of Batada {\it et al.}~\cite{batada}. The latter is the most 
recent. It was obtained by extracting the most reliable interactions
from a vast data base which includes the data of both Bu {\it et al.} and 
Krogan {\it et al.}. In Fig.~\ref{ratios-compare} we plot the ratios
between the actual counts and the average counts in rewired networks
for Bu {\it et al.} and for the HC data set against the analogous 
ratios for the Krogan {\it et al.} networks. If the three data sets
indeed describe the same yeast network -- as they purport to do, within 
experimental uncertainties -- the points should all fall onto the 
diagonal. Instead, we see systematic deviations. Surprisingly, these 
deviations are much stronger between the Krogan {\it et al.} and the 
HC networks than between the Krogan {\it et al.} and the Bu {\it et al.} 
networks. Clarifying these and other systematic irregularities should 
give valuable insight into the strengths and weaknesses of the methods 
used in constructing the networks as well as their biological 
reliability, and should lead to improved methods for network 
reconstruction.

In the present paper we have only dealt with undirected networks. The
basic sampling algorithm works equally well for directed networks. The
main obstacle in applying our methods to the latter is the huge number
of directed subgraphs, even for relatively small sizes.
Nevertheless, we will present an analysis of directed networks in
forthcoming work, as well as applications to other undirected
networks.

Acknowledgements: We thank Gabriel Musso for valuable information on 
the yeast network.


\begin{thebibliography}{99}
\bibitem{footnote} The networks given in \cite{ecoli,yeast} are not
   connected. In the present paper we used only their largest connected
   components.
\bibitem{footnote2} We might try to estimate the optimal $p$ by the
   threshold for an infinite SIR epidemic on an infinite tree like
   network with the same degree distribution, $p_c = \langle k\rangle
   / \langle k^2\rangle$ \cite{pastor}. For the two networks
   considered in this paper, this would give $p_c({\rm yeast}) =
   0.062$, $p_c({\rm e.~coli}) = 0.070$, i.e.  a much smaller
   difference in the optimal $p$ values. One reason why this is not
   observed might be the very strong clustering, in particular in the
   yeast data, which is neglected in this argument.
\bibitem{footnote3} Another reason why no minimum appears in the
   $n=8$ curve is that we kept the number of generated clusters fixed,
   not CPU time.  Since larger $p$ values also imply larger clusters
   in average, the CPU time per cluster increases sharply for larger
   $p$.
\end{thebibliography}
\end{document}